\documentclass[aps,prl,twocolumn,naturefem,amsmath,amssymb]{revtex4}
\usepackage{graphicx}
\usepackage{amsmath}
\usepackage{bm}
\usepackage{times}
\usepackage{bm}
\usepackage{amsmath,amsthm,amsfonts,amssymb,amscd,color}

\newcommand{\ket}[1]{\ensuremath{|{#1}\rangle}} 

\begin{document}

\title{Quantum-Enhanced Sensing from Hyper-Entanglement}
\author{S. P. Walborn}
\affiliation{Instituto de F\'isica, Universidade Federal do Rio de Janeiro, P.O.Box 68528, Rio de Janeiro, RJ 21941-972, Brazil}
\author{A. H. Pimentel}
\affiliation{Instituto de F\'isica, Universidade Federal do Rio de Janeiro, P.O.Box 68528, Rio de Janeiro, RJ 21941-972, Brazil}
\author{L. Davidovich}
\affiliation{Instituto de F\'isica, Universidade Federal do Rio de Janeiro, P.O.Box 68528, Rio de Janeiro, RJ 21941-972, Brazil}
\author{R. L. de Matos Filho}
\affiliation{Instituto de F\'isica, Universidade Federal do Rio de Janeiro, P.O.Box 68528, Rio de Janeiro, RJ 21941-972, Brazil}
\begin{abstract}
Hyperentanglement --- simultaneous entanglement between multiple degrees of freedom of two or more systems --- has been used to enhance quantum information tasks such as quantum communication and photonic quantum computing.  Here we show that hyperentanglement can lead to increased quantum advantage in metrology, with contributions from the entanglement in each degree of freedom, allowing for Heisenberg scaling in the precision of parameter estimation.  Our experiment employs photon pairs entangled in polarization and spatial degrees of freedom to estimate a small tilt angle of a mirror.   Precision limits beyond shot noise are saturated through a simple binary measurement of the polarization state. The broad validity of the dynamics considered here implies that similar strategies based on hyperentanglement can offer improvement in a wide variety of metrological tasks.    
 \end{abstract}
\pacs{03.65.Ta, 03.67.Ac, 42.50.Lc, 06.20.-f}
\maketitle

 To exploit the advantage of quantum entanglement for tasks such as computation and metrology requires producing high-dimensional quantum states composed of many entangled sub-systems \cite{noon,dowling,qmetro1,qmetro2}.  In addition to the difficulty in producing large entangled states, they are typically very sensitive to noise \cite{aolita1,walmsley1,aolita2}.  An alternative route to enhance the size of  a system is through hyperentanglement \cite{kwiat97}, that is, the entanglement in multiple degrees of freedom (DOF) of a composite quantum system.  So far,  hyperentanglement has found use in high-capacity quantum communication \cite{kwiat98a,walborn03c,almeida06,barreiro08,graham15}, photonic quantum computing \cite{walborn05c, vallone10}, tests of quantum non-locality \cite{cinelli05,barbieri06}, and the direct characterization of entanglement \cite{steve} and quantum dynamics \cite{graham13}.  Here we demonstrate the usefulness of hyperentanglement in metrology, allowing one to reach the ultimate quantum precision limits, for the paradigmatic case of an interaction between two degrees of freedom of the same system.  Examples of this type of evolution include the interaction between spin and momentum in a Stern-Gerlach experiment, between internal and external DOF of trapped ions \cite{rmpion}, or the polarization and spatial DOF of an optical field propagating through a birefringent medium \cite{biref}.  
 \par
  Our experiment employs hyperentangled photon pairs to monitor a tiny rotation of a mirror. Entanglement in both spatial and polarization DOF leads to increased quantum advantage in metrological sensing, allowing for Heisenberg scaling in the number of photons $N$ used to probe the rotation: the precision of estimation becomes proportional to $1/N$, instead of the shot-noise behavior $\propto 1/\sqrt{N}$.    
Estimation of tilt angles of mirrors is important in several fields of science \cite{kara1981,gillies,virgo,atom}. A common procedure consists in detecting the  spatial or phase displacement of a laser beam reflected by the mirror \cite{park, howell,beamdeflection,alves2014weak,unbalancedweak,lyons16, alves2017}.  Our method, on the other hand, is based solely on a binary polarization measurement at the output of an interferometer with a displaced input beam.  This scheme takes advantage of both hyper-entanglement and beam displacement to increase the precision  beyond the shot-noise limit.  
 \par
{\it Precision measurements and Fisher information.}
\begin{figure}
\begin{center}
\includegraphics[width=8cm]{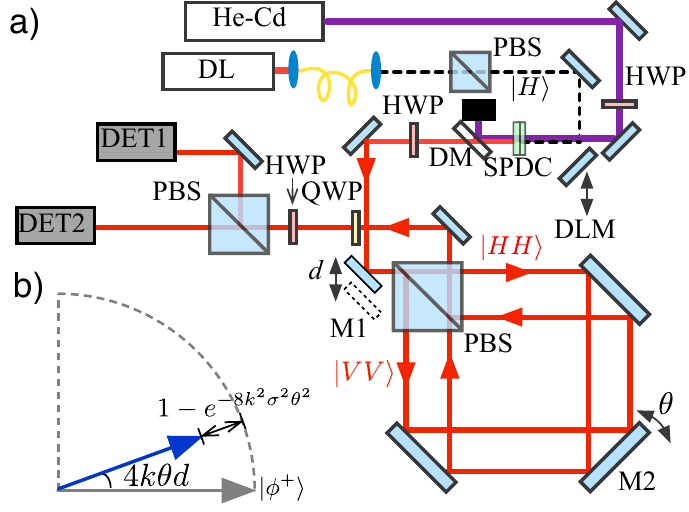}
\end{center}
\caption{\label{fig:setup}  a) Experimental setup (see text).  b) One quadrant of the equatorial plane of the Bloch sphere. Tilt of mirror M2 causes the Bloch vector of the initial polarization state $\ket{\phi^+}$ to shrink and rotate.} 
\end{figure}
Bounds on the uncertainty in the estimation of a parameter $\theta$  can be assessed by the Cram\'er-Rao inequality \cite{cramer}:
\begin{equation}\label{cr1}
\delta\theta\ge{1\over\sqrt{\nu F(\theta)}}\,,
\end{equation} 
where 
\begin{equation}\label{fisher}
F(\theta)=\sum_j\frac{1}{p_j(\theta)}\left[{dp_j(\theta)\over d\theta}\right]^2
\end{equation}
 is the Fisher information on $\theta$, for a given measurement on the probe used to estimate the parameter,  $p_j(\theta)$ is the probability of obtaining experimental result $j$, given that the value of the parameter is $\theta$, and $\nu$ is the number of repetitions of the measurement. Maximization of $F(\theta)$ over all possible quantum measurements yields the quantum Fisher information 
${\cal F}(\theta)$, which leads to the ultimate precision bound. Since inequality \eqref{cr1} can be saturated in the limit of large $\nu$,  the Fisher information can be considered as a figure of merit for the precision of a given measurement strategy.  We adopt this standpoint in the following.


\par
 {\it A hyperentanglement-enhanced tilt sensor.}
FIG. \ref{fig:setup} shows our experimental setup. A 325nm He-Cd laser pumps two BBO crystals (2mm length), producing polarization and spatially entangled photon pairs via colinear spontaneous parametric down-conversion (SPDC) \cite{kwiat99,walborn10} with degenerate wavelength 650 nm.  The state of the twin photons is well-described by $\ket{\Psi^{(2)}}=\ket{\Phi^+}\otimes\ket{\psi^{ (2)}}$, where  $\ket{\psi^{(2)}}$ stands for the spatial degrees of freedom of the photon pair, and $\ket{\Phi^+}=(\ket{HH}+\ket{VV})/\sqrt{2}$ is the polarization state, with $H$ ($V$) standing for horizontal (vertical) polarization.  In usual conditions, photons pairs from SPDC display a large amount of spatial entanglement, which appears in the form of  correlations of the photons in the near-field, and anti-correlations in the far-field \cite{walborn10} (see Appendix I).    A pair of lenses (not shown) is used to map the source plane onto mirror M2.    
\par
The pump beam is removed via a dichroic mirror (DM).  Mirror M1 is mounted on a translation stage to adjust the initial beam displacement $d$. Both photons are sent into a Sagnac interferometer built with a polarizing beam splitter (PBS).  The $H$-polarized ($V$-polarized) photons propagate in the clockwise (counter-clockwise) direction.  When $d \neq 0$, the two trajectories are spatially displaced in the transverse plane, as shown in the sketch of the experiment.
   Small angular deflections $\theta$ of mirror M2 are controlled via a stepper motor and a piezo-electric actuator.     
\par
A half-wave plate (HWP) and quarter-wave plate (QWP) in conjunction with a PBS are used to project onto different polarization states.   Photons are detected in coincidence, each single-photon detector equipped with an 8mm diameter circular aperture and a 10nm bandwidth filter.  For measurements with independent photons, the mirror DLM is placed in the setup, coupling light from a 650nm diode laser (DL) into the interferometer.  These photons are coupled out of a single-mode fiber and prepared in the polarization state $\ket{\phi^+}=(\ket{H}+\ket{V})/\sqrt{2} $ by a HWP and are aligned to follow the same propagation path as the entangled photons, with the spatial state $|\psi^{(1)}\rangle$ corresponding to a Gaussian beam profile, so that the total state is $|\Psi^{(1)}\rangle=|\phi^+\rangle\otimes|\psi^{(1)}\rangle$. 
\par
Tilt of mirror M2 induces transverse momentum shifts in opposite directions on the two counter-propagating beams inside the interferometer, producing entanglement between the polarization and spatial DOF of each photon.    Moreover, the tilt also induces a relative phase between the two counter-propagating beams that is proportional to the displacement $d$, due to the different path lengths inside the interferometer.  Both of these effects imprint information about $\theta$ in the polarization state of the photons. For single photons, entanglement between the two DOF leads to purity loss of the polarization state, which shrinks the length of the Bloch vector from unity to $\exp(-8k^2\sigma^2\theta^2)$, where $k$ is the photon wave number and $\sigma$ is the width of the transverse profile of the Gaussian beam at mirror M2. The relative phase between the $H$ and $V$ polarization components leads to a rotation of the Bloch vector around the $z$ axis, by an angle $4\theta kd$.  Both of these effects are illustrated in FIG.~\ref{fig:setup}b). A similar analysis applies to the two-photon state $\ket{\Phi^+}$, which here behaves as an effective two-level system. However, in this case rotation and reduction of the Bloch vector can be up to two times larger, leading to increased precision in the estimation of $\theta$.
\par
{\it Heisenberg scaling via hyper-entanglement.}
The action of the interferometer on a single photon can be represented by the unitary operator:
\begin{equation}\label{unitary}
\hat U^{(1)}(\theta)  = e^{-i2k\theta\hat\sigma_z\hat{x}},
\end{equation}
where $\hat\sigma_z=|H\rangle\langle H|-|V\rangle\langle V|$, $\hat{x}$ represents the transverse position operator, and $2k\theta$ is the shift in transverse momentum.   For two photons, one has instead $\hat U^{(2)}(\theta)=\exp[-i2k\theta(\hat\sigma_{z_1}\hat x_1+\hat\sigma_{z_2}\hat x_2)]$. 
\par
For a unitary evolution $\hat U(\theta)$ of a pure state, the quantum Fisher information ${\cal F}(\theta)$ is given by four times the variance in the initial state of the generator of $\hat U(\theta)$ (see Appendix II). 
From Eq. \eqref{unitary}, the quantum Fisher information for single photons is then
\begin{equation}\label{single}
{\cal F}^{(1)}(\theta)=16k^2\left(\sigma^2+d^2\right),
\end{equation}
where $d= \langle \hat{x} \rangle$ is the initial transverse displacement of the beam. The terms proportional to $\sigma^2$ originates from the momentum shift.  The contribution $d^2$ arises from the relative path difference between the $H$ and $V$ polarization components, and increases ${\cal F}(\theta)$ with respect to the balanced situation $d=0$. If $d\gg\sigma$, this can lead to a huge increase in the precision of estimation of $\theta$. In this case, the information on $\theta$ is contained mainly in the relative phase.  Though this effect has been exploited by a few authors \cite{park, howell}, it is not present in  implementations using colinear Sagnac configurations \cite{beamdeflection,alves2014weak,unbalancedweak,alves2017}.  Furthermore, we rely in the present scheme on a simple binary polarization measurement.
\par
The quantum Fisher information for entangled photon pairs unveils subtle effects related to both the polarization and spatial entanglement. 
Assuming the usual spatial state of the SPDC photons \cite{walborn10}, one  has
\begin{equation}\label{double}
{\cal F}^{(2)}(\theta)=32k^2\left(\sigma^2+\mathrm{Cov}[x_1,x_2]+2d^2\right)\,,
\end{equation}
where $\sigma^2$ is the variance of the spatial profile of each photon, and the covariance $\mathrm{Cov}[ x_1, x_2]$ corresponds to the transverse spatial correlations between the photon pairs.

There are remarkable differences between Eqs.~\eqref{single} and \eqref{double}. The overall multiplicative factor for  photon pairs is twice that of  single photons, since there are two photons passing through the interferometer.  The extra factor of two multiplying $d^2$ originates entirely from the polarization entanglement, which is thus seen to enhance the precision of estimation of $\theta$. 
The other two terms in \eqref{double} stem from the momentum shift.   If $\mathrm{Cov}[x_1,x_2]=0$, there is no gain with respect to using two spatially independent photons. On the other hand, the maximum value of the covariance is $\sigma^2$, which corresponds to perfect spatial correlations. In this limiting case, the quantum Fisher information is maximum: 
\begin{equation}\label{Heisenberg}
{\cal F}^{(2)}_ {\rm max}(\theta)=64k^2\left(\sigma^2+d^2\right),
\end{equation}
which displays a dependence with the square of the number of photons, when compared to Eq. \eqref{single}. This is the well-known Heisenberg scaling for $N$ systems:  the Fisher information is proportional to $N^2$, rather then to $N$, which corresponds to the shot-noise behavior (two independent photons). This enhanced scaling is due to the joint contribution of maximal spatial correlation and maximal polarization entanglement. Heisenberg scaling can also be demonstrated for $N$ maximally entangled photons, as we show in Appendix II.
\par
The quantum Fisher information $\mathcal{F}$ provides the ultimate precision limit.  However, it can be challenging to find an experimentally friendly measurement strategy that allows one to reach these bounds.   An advantage of the present scheme is that in the interesting limit of very small tilt angles, $(\theta k\sigma)^2\ll 1$ and $(\theta kd)^2\ll1$,  the quantum Fisher information can be reached by projecting the final polarization state onto the basis defined by the intial state and the one orthogonal to it. 
\par
Applying $\hat U^{(\ell)}(\theta)$ to the initial state, it is straightforward to show that the probability $p^{(\ell)}(\theta)$ to project the final polarization state onto the initial state is
\begin{equation}\label{pf}
p^{(\ell)}(\theta)=\frac{1}{2}\left[1+ V\cos(4\ell \theta k d)\exp\left(-8\ell \theta^2 k^2\sigma_{(\ell)}^2\right)\right],
\end{equation}
where $\ell=1$ ( $\ell=2$)  for single (entangled) photons, 
$\sigma_{(\ell)}^2\equiv \sigma^2+(\ell-1)\mathrm{Cov}[x_1,x_2]$ and $V$ is the visibility of the interferometer.  The corresponding Fisher informations are given by
\begin{equation}\label{fisher2}
F^{(\ell)}(\theta)=\frac{16V^2\ell^2k^2\left[ d\sin(4\ell \theta kd)+4 k\theta\sigma_{(\ell)}^2\cos(4\ell \theta kd)\right]^2}{\exp\left(16\ell \theta^2k^2\sigma_{(\ell)}^2\right)-V^2\cos^2(4\ell \theta kd)}.
\end{equation}
When $\left(\theta k\sigma_{(\ell)}\right)^2\ll 1$, $(\theta kd)^2\ll1$, and  $V=1$, this expression coincides with \eqref{single} for $\ell=1$ and with \eqref{double} for $\ell=2$, up to terms of second order in $\theta k\sigma_{(\ell)}$ and $\theta kd$. Thus,  within this approximation, the binary polarization measurement reaches the maximum precision limit given by the quantum Fisher information in both the single- and the  two-photon case.   
\par
{\it Experimental demonstration of quantum enhancement.}
The hyperentangled photons were injected into the Sagnac interferometer, and the mirror angle $\theta$ was scanned.  We measured the probability $p^{(2)}(\theta)$ to project the two-photon state at the output of the interferometer onto the polarization state $\ket{\Phi^+}$, as a function of $\theta$.   For the experiment with independent photons, the probability $p^{(1)}(\theta)$ to detect  the state $\ket{\phi^+}$ was obtained as a function of $\theta$.   In both cases, we controlled the initial displacement $d$ of the photon beams by translating their initial positions using mirror M1. 
\par
\begin{figure}
\begin{center}
\includegraphics[width=8cm]{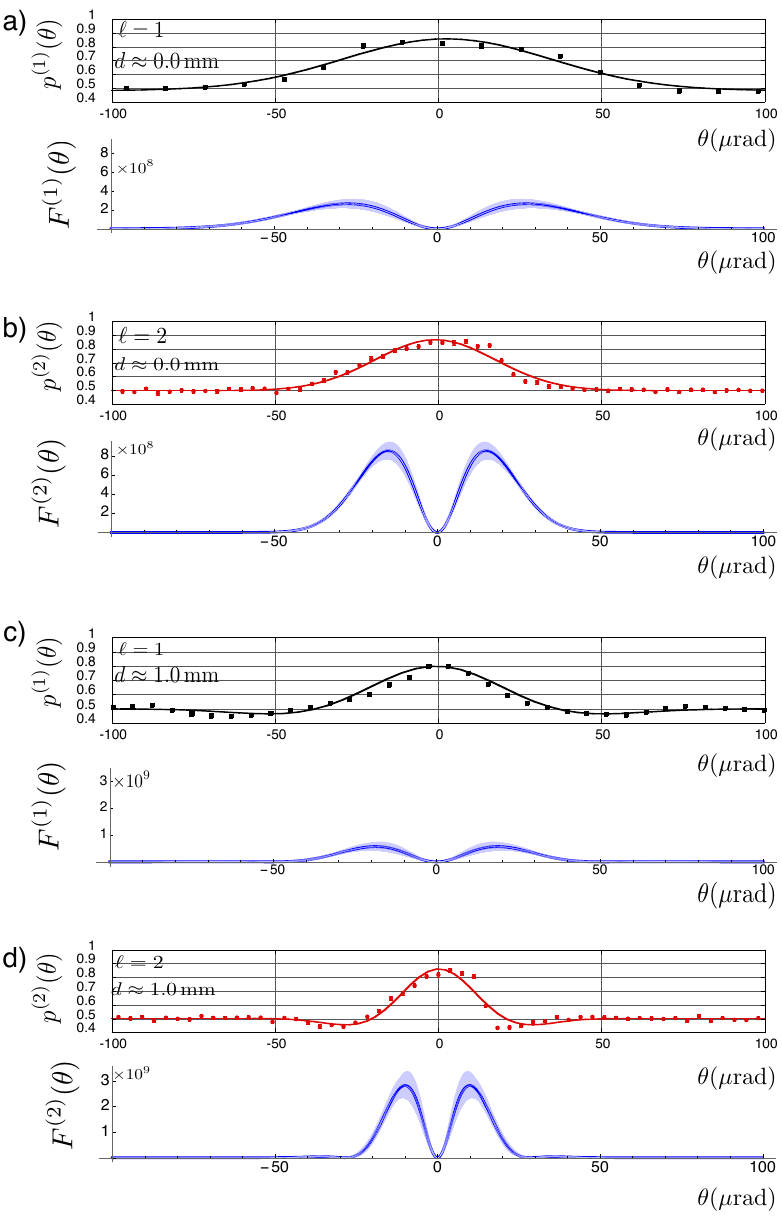}
\end{center}
\caption{\label{fig:Phi-d} Experimental probability $p^{(\ell)}(\theta)$ and corresponding Fisher Information $F^{(\ell)}(\theta)$ (blue curves) as a function of $\theta$ for different values of the displacement $d$. Black squares correspond to independent photons ($\ell=1$),  and the red circles to hyperentangled photon pairs ($\ell=2$). The red and black curves are fits to Eq. \eqref{pf}.  From the fit parameters we plotted the theoretical curves (in blue) for the Fisher Information, given by Eq.~\eqref{fisher}.} 
\end{figure}
FIG. \ref{fig:Phi-d} compares experimental results for $p^{(\ell)}(\theta)$ obtained for independent photons (plots a) and c)) and entangled photon pairs (plots b) and d)).  The red circles and black squares are data points and the solid curves are fits using Eq. \eqref{pf}.  Error bars were calculated from propagation of Poissionian count statistics, and are smaller than the plot symbols in most cases.  From the curve fits we extract the visibility $V$, as well as the variances $\sigma_{(\ell)}^2$.    The fits return values $\sigma_{(1)}^2 \approx 0.65 \pm 0.05 \mathrm{mm}^2$  and $\sigma_{(2)}^2 \approx 1.1 - 1.34\, \mathrm{mm}^2$,  with typical uncertainty less than $0.1\, \mathrm{mm}^2$. These agree with independent measurements of the single and two-photon beam near mirror M2, obtained by scanning a thin slit aperture, giving $\sigma_{(1)}^2=0.65 \pm 0.02\mathrm{mm}^2$ and $\sigma_{(2)}^2=1.22\pm0.04\,\mathrm{mm}^2$.    The variance of the individual  beams of the photon pair at mirror M2 were both $\sigma^2 = 0.70 \pm 0.03 \mathrm{mm}^2$.   The displacement $d$ was left as a free fit parameter, and estimates returned values comparable to the expected ones.  Using the values obtained from the curve fits to $p^{(\ell)}(\theta)$ we plot the corresponding Fisher informations using Eq.~\eqref{fisher2}, given by the blue curves, with the experimental uncertainty given by the light shaded regions.  

In the plots shown in FIG.~\ref{fig:Phi-d} one can see the damped oscillatory behavior of $p^{(\ell)}(\theta)$,  where the oscillation frequency is proportional to the beam displacement $d$.   We obtained visibilities in the range $V \approx 0.72-0.77 \pm 0.01$ for the two-photon case, which is much less than the ideal case $V=1$.  To exploit the full range of spatial entanglement, the entire cross section of the two-photon must be detected. This reduces $V$, since the relative phase of the two-photon polarization state varies along the transverse direction of the two-photon beam \cite{kwiat99}.   The visibilities could be improved in future experiments by using thinner non-linear crystals, narrower spectral filters, and/or additional compensation crystals \cite{altepeter05}.  The independent-photon experiment leads to higher visibilities. However, in order to have a direct comparison between the two cases, in the independent-photon experiment we reduce the visibility by slightly misaligning the interferometer. Comparison of plot \ref{fig:Phi-d} a) with b), and c) with d),  clearly displays the quantum enhancement: the maximum Fisher information corresponding to the photon pair is approximately four times the corresponding one for single photons. This shows that hyperentanglement leads to an increase in the precision of estimation, as compared to the independent-photon situation. It is important to stress that the displacement $d$ fulfills  an important role in increasing the Fisher information in both cases. \par
\begin{figure}
\begin{center}
\includegraphics[width=8cm]{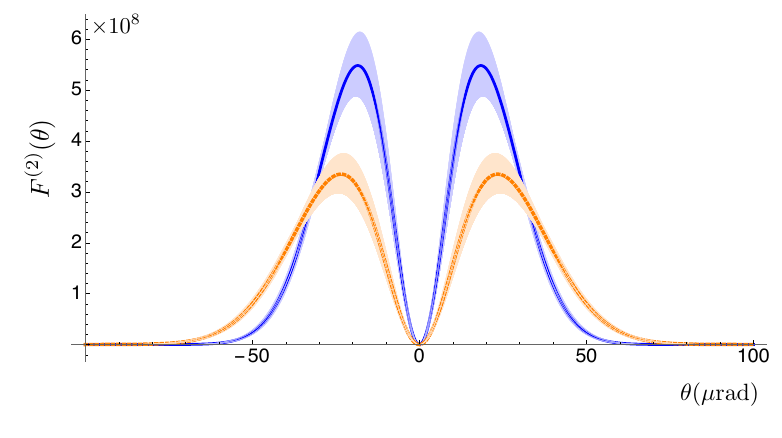}
\end{center}
\caption{\label{fig:Phi-Ent-Sep} Experimentally determined Fisher Information $F^{(2)}(\theta)$ as a function of $\theta$ for the input state $\ket{\Phi^+}$ for different values of the initial spatial correlation, and  $d=0$.  The blue solid curve corresponds to larger correlation $C=0.84 \pm0.14$, while the orange dashed curve corresponds to $C=0.18\pm0.10$ (see text).} 
\end{figure}

To show that there is indeed an increased sensitivity due to the correlations coming from spatial entanglement, we used a spatial filter to alter the spatial correlation between the photons, in such a way that the widths of the marginal distributions were kept constant.  FIG. \ref{fig:Phi-Ent-Sep} shows the resulting Fisher informations for $d=0$, when only the effect of the momentum shift is relevant.  Let us quantify the spatial correlation using a correlation coefficient $C = \mathrm{Cov}[x_1,x_2]/\sigma^2$. The blue solid curves correspond to high spatial correlation $C=0.84 \pm0.14$, while the orange dashed curves to the case where $C=0.18\pm0.10$.   The plots clearly display the contribution from spatial entanglement to the Fisher information.  

\begin{figure}
\begin{center}
\includegraphics[width=8cm]{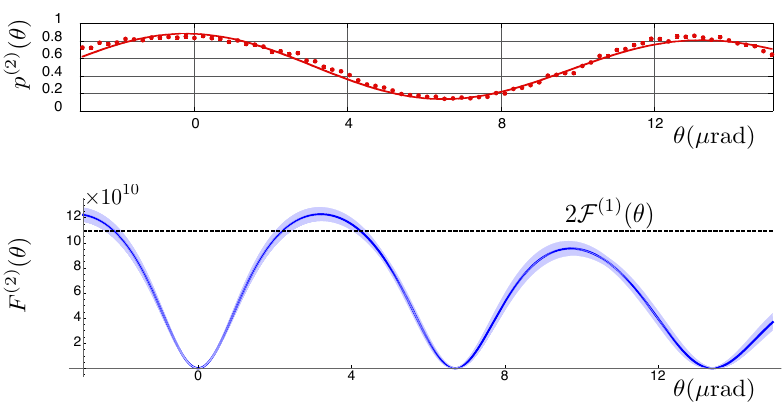}
\end{center}
\caption{\label{fig:Phi-dLarge}  Experimental probability $p^{(2)}(\phi^+)$ (red circles) and corresponding Fisher Information $F_2$ (blue curve) as a function of $\theta$ for  $d \approx 6$mm.  The red curve is a fit using Eq. \eqref{pf} for $\ell=2$.   From the fit parameters the Fisher Information is plotted, using Eq.~\eqref{fisher}. Precision beyond shot-noise, given by the horizontal dashed line, is clearly displayed.} 
\end{figure}

Even with a non-ideal visibility,  $V=0.77\pm0.01$, it was possible to achieve sub-shot noise precision, as displayed in FIG. \ref{fig:Phi-dLarge}, where $d= 5.97\pm0.05$ mm.  The enhancement in the Fisher information for $d >> \sigma$ is due primarily to polarization entanglement.   
 The dashed line corresponds to the quantum Fisher information in the shot noise limit.  It represents the maximal information about $\theta$ that can be retrieved by sending two independent photons into the interferometer, in the ideal situation of $V=1$.       It is also important to note that the additivity of the Fisher information implies that $N/2$ photon pairs in our scheme ($N$ photons total) beat the shot-noise limit for $N$ independent photons, given by $N \mathcal{F}^{(1)}(\theta)$.  
\par
We note from FIG.~\ref{fig:Phi-d}   and  FIG.~\ref{fig:Phi-dLarge} that the sub-shot noise regime is achieved only for larger values of $d$. Analyzing Eq.~\eqref{fisher2} for $d=0$ and $d \gg \sigma$, we see that this is due to the fact that the momentum shift contribution is much more  sensitive to the loss of visibility, as compared to the rotation due to the path difference between the two polarization components. This last contribution becomes more important as $d$ increases, thus mitigating the effect of visibility loss.  

\par

\par
 {\it Summary and outlook.}
We have demonstrated that hyperentanglement can be an important and useful resource for quantum metrology, allowing for  the attainment of Heisenberg scaling and the sub-shot noise regime, even in the presence of experimental imperfections.   Our method uses a displaced input beam in a Sagnac interferometer to greatly boost the precision in the estimation of the tilt angle of a mirror. This allows for multiple degrees of freedom to contribute, leading to increased sensitivity even for independent photons, when compared to schemes employing colinear Sagnac configurations. Furthermore, this displacement is essential to exploit the quantum gain in sensitivity that results from polarization entanglement. 
\par
The scheme presented in this paper is particularly simple and reliable, since it involves only a binary polarization measurement. Furthermore, the source for polarization entanglement already yields the necessary spatial entanglement, via conservation of momentum.   Future improvements, such as increasing the visibility $V$ by using thinner non-linear crystals, narrower spectral filters, and/or  compensation crystals, should lead to a useful metrological tool. The ideas developed here have a broad range of applications, since the basic dynamics, involving the bilinear coupling between a continuous and a discrete bi-dimensional degrees of freedom, are common to many other systems.    

\begin{acknowledgments}
The authors acknowledge financial support from the Brazilian funding agencies CNPq, CAPES and FAPERJ, and the National Institute of Science and Technology for Quantum Information.
\end{acknowledgments}

\appendix
\section{Appendix I: Spatial entanglement from SPDC}
Within the monochromatic and paraxial approximations, the spatial state of the photon pair at the near-field of the BBO source can be written as \cite{walborn10}
\begin{equation}
\ket{\psi^{(2)}} = \iint d\mathbf{q}_1 d\mathbf{q}_2 \mathcal{W}(\mathbf{q}_1+\mathbf{q}_2) \Gamma(\mathbf{q}_1-\mathbf{q}_2) \ket{\mathbf{q}_1}\ket{\mathbf{q}_2},
\label{eq:psi}
\end{equation}
where $\ket{\mathbf{q}_1},\ket{\mathbf{q}_2}$ are transverse-position eigenstates, $\mathcal{W}$ is the Gaussian transverse spatial profile of the pump laser beam and $\Gamma$ is the phase matching function, and we ignore effects due to birefringence of the crystals.  In usual conditions, the state \eqref{eq:psi} can display a large amount of entanglement, which appears in the form of  spatial correlations of the photons in the near-field, and anti-correlations in the far-field \cite{walborn10}.  Moreover, these correlations can be observed simultaneously with the polarization correlations \cite{almeida06}.        

\section{Appendix II: Quantum Fisher Information for $N$ hyperentangled probe systems}
The Quantum Fisher information for the estimation of the tilt angle $\theta$, for $N$-partite pure states under unitary evolution is $\mathcal{F}^{(N)}(\theta) = 4 \langle \Delta \hat{H}^2 \rangle$, where in the present case the Hamiltonian $\hat{H}$ is given by $2k\sum_{j=1}^{N} \hat{\sigma}_{z_j} \hat{x}_j$.  Consider an initial $N$-partite pure state, where each physical system contains one two-level degree of freedom and one continuous degree of freedom.  Let the $N$-partite hyperentangled state be $\ket{S} = \ket{\Phi}\otimes\ket{\psi}$, where $\ket{\Phi}=(\ket{0}^{\otimes N}+\ket{1}^{\otimes N})/\sqrt{2}$ is a GHZ state of $N$ qubits and $\ket{\psi}$ refers to a state of  $N$ quantum continuous variables.  Assuming, as was done in relation to Eq. (6) of the main text, that for each system we have  $\langle \Delta \hat{x}_j \rangle = \sigma^2$, and $\langle \hat{x}_j\rangle=d$, it is straightforward to show that  
\begin{equation}\label{fn}
\mathcal{F}^{(N)}(\theta) = 16k^2\big(N\sigma^2+N^2d^2+ 2 \sum_{j=1}^{N} \sum_{i=1}^{j-1} {\rm Cov}[{x}_i, {x}_j]\big), 
\end{equation}   
where we used the fact that $\langle \hat{\sigma}_{z_j} \rangle=0$ and $\langle \hat{\sigma}_{z_j} \hat{\sigma}_{z_i} \rangle=1$.  Equation \eqref{fn} generalizes Eq. (6) of the text.  In the limit in which the covariances attain their maximum value, equal to $\sigma^2$, one has 
\begin{equation}\label{fnmax}
\mathcal{F}_{\rm max}^{(N)}(\theta) = N^216k^2\left(\sigma^2+ d^2\right)=N^2{\cal F}^{(1)}(\theta)\,,
\end{equation}
which clearly displays Heisenberg scaling. 

In our experiment, $N=2$, $\ket{\psi^{(2)}}$ refers to the spatial degrees of freedom of twin photons, given by Eq. (3), while $\ket{\Phi}=\ket{\Phi^+}$ is a maximally entangled polarization state. In this case, Eq.~\eqref{fn} leads to Eq.~(6), and Eq.~\eqref{fnmax} to Eq.~(7).


\end{document}